\documentclass[journal]{IEEEtran}
\ifCLASSINFOpdf
\else
\fi
\usepackage{tabularx}
\usepackage{makecell}
\usepackage{graphicx}
\usepackage{subfig}
\usepackage[dvipsnames]{xcolor}
\usepackage[pagewise,switch]{lineno}
\usepackage{caption}
\usepackage[font=footnotesize]{caption}
\usepackage{amsmath}
\usepackage [autostyle, english = american]{csquotes}
\MakeOuterQuote{"}
\usepackage{hyperref}

\begin{document}
%
\title{Using Eye-tracking Data to Predict Situation Awareness in Real Time during Takeover Transitions in Conditionally Automated Driving}
%
%
%

\author{{Feng Zhou, X. Jessie Yang, and Joost de Winter}
\thanks{Corresponding author: Feng Zhou.}
\thanks{F. Zhou is with the Department of Industrial and Manufacturing Systems Engineering, University of Michigan Dearborn, 4901 Evergreen Rd. Dearborn, MI 48128 USA (e-mail: fezhou@umich.edu).}
\thanks{X. J. Yang is with the Department of Industrial and Operations Engineering, University of Michigan, Ann Arbor, 1205 Beal Avenue, Ann Arbor, MI 48109 USA (e-mail: xijyang@umich.edu).}
\thanks{J. de Winter is with the Cognitive Robotics Department, Faculty of Mechanical, Maritime and Materials Engineering, Delft University of Technology, Mekelweg 2, 2628 CD Delft, the Netherlands (e-mail: J.C.F.deWinter@tudelft.nl).}
\thanks{Manuscript received on December 16, 2020, revised on Feb. 1, 2021, and accepted on March 25, 2020.}}

%
%

\markboth{Accepted by IEEE Transactions on Intelligent Transportation Systems, 2021}%
{Zhou \MakeLowercase{\textit{et al.}}: Using Eye-tracking Data to Predict Situation Awareness in Real Time during Takeover Transitions in Conditionally Automated Driving}
%



\maketitle

\begin{abstract}
Situation awareness (SA) is critical to improving takeover performance during the transition period from automated driving to manual driving. Although many studies measured SA during or after the driving task, few studies have attempted to predict SA in real time in automated driving. In this work, we propose to predict SA during the takeover transition period in conditionally automated driving using eye-tracking and self-reported data. First, a tree ensemble machine learning model, named LightGBM (Light Gradient Boosting Machine), was used to predict SA. Second, in order to understand what factors influenced SA and how, SHAP (SHapley Additive exPlanations) values of individual predictor variables in the LightGBM model were calculated. These SHAP values explained the prediction model by identifying the most important factors and their effects on SA, which further improved the model performance of LightGBM through feature selection. We standardized SA between 0 and 1 by aggregating three performance measures (i.e., placement, distance, and speed estimation of vehicles with regard to the ego-vehicle) of SA in recreating simulated driving scenarios, after 33 participants viewed 32 videos with six lengths between 1 and 20 s. Using only eye-tracking data, our proposed model outperformed other selected machine learning models, having a root-mean-squared error (RMSE) of 0.121, a mean absolute error (MAE) of 0.096, and a 0.719 correlation coefficient between the predicted SA and the ground truth. The code is available at \url{https://github.com/refengchou/Situation-awareness-prediction}. Our proposed model provided important implications on how to monitor and predict SA in real time in automated driving using eye-tracking data.
\end{abstract}


%
\IEEEpeerreviewmaketitle

\section{Introduction}
%
%
%
%
\IEEEPARstart {C}{onditionally} automated vehicles (i.e., SAE Level 3 \cite{sae2018taxonomy}) have the potential to improve driving safety and mobility while allowing drivers to conduct non-driving related tasks (NDRTs) \cite{AYOUB2021102}. However, as drivers are involved in NDRTs, they can be out of the control loop for a prolonged duration \cite{braunagel2015driver}, which can significantly reduce their situation awareness (SA) \cite{endsley1995toward} required to successfully take over control of the automated vehicle when it reaches its system limit \cite{ayoub2019manual,du2020examining,zhou2019takeover,du2020predicting}. For example, Zeeb et al. \cite{zeeb2016take} found that, compared to those who were not performing NDRTs, drivers performing NDRTs before the takeover request exhibited significantly deteriorated takeover performance, even though they achieved motor readiness quickly. Braunagel et al. \cite{braunagel2015driver,braunagel2017online} argued that automated driving systems should be able to recognize NDRTs (using eye-tracking data with machine learning models) during conditionally automated driving in order to measure and monitor driver's SA during the takeover transition period. With such prediction and monitoring systems, appropriate interventions can be introduced to improve takeover performance \cite{du2020predicting,du2020predictingchi}. 

Endsley \cite{endsley1995toward} defined SA as consisting of three levels, namely 1) the perception of environmental elements and events in time or space, 2) the comprehension of their meaning, and 3) the projection of their status in the future. SA is essential to ensure a successful takeover transition in conditionally automated driving. However, little research has attempted to monitor and predict SA in real time. Existing studies mainly used standardized tools, such as the situation awareness global assessment technique (SAGAT) \cite{endsley1995measurement} and the situation awareness rating technique (SART) \cite{taylor2017situational} to measure SA. While SAGAT is a performance-based tool that asks participants to answer questions about the three levels of SA \cite{endsley1995toward}, it faces operational challenges in real applications due to its freeze-probe nature, especially in time-sensitive scenarios such as takeovers. As a result, SAGAT has rarely been used in naturalistic driving \cite{sirkin2017toward}. As a typical example of a self-report tool, SART measures the amount of demand on attentional resources, the supply of attentional resources, and the understanding of the situation to obtain one's SA (i.e., SA = understanding – (demand – supply)) \cite{taylor2017situational}. SART is usually administered during or after the task is finished. For example, Petersen et al. \cite{petersen2019situational} used the SART tool after the task to measure drivers' SA in automated driving and found that high SA increased trust in automated driving and yielded improved performance of NDRTs. However, similar to SAGAT, SART cannot be used to measure SA in real time without interfering with the task at hand. Given the importance of SA prior to or during the takeover transition, it is important to assess SA in real time using unobtrusive measures. 

Recently, researchers have used eye-tracking data to assess SA dynamically. Eye-tracking measures are indicators of visual attention, which in turn is of critical importance to perceiving, comprehending, and projecting the unfolding takeover process. For example, Young et al. \cite{young2013missing} and Molnar \cite{molnar2017age} found that the time percentages of eyes-on-the-road were related to drivers' SA in naturalistic manual driving and simulated automated driving, respectively. Yoon and Ji \cite{yoon2019non} found that eye-tracking measures, such as the time needed for drivers to shift their attention from NDRTs and fixate on the road, played an important role in re-engaging the driving task during the takeover process. De Winter et al. \cite{de2019situation} found that visual-sampling scores obtained using an eye tracker correlated more strongly with task performance than freeze-probe scores acquired via a SAGAT-like method. However, these studies utilized post-analysis methods and few studies attempted to predict SA using eye-tracking data in real time. 

In this study, we proposed a machine learning model to predict SA in takeover transitions based on eye-tracking data using both LightGBM (Light Gradient Boosting Machine) \cite{ke2017lightgbm} and SHAP (SHapley Additive exPlanations) \cite{lundberg2018explainable,lundberg2020local}. First, LightGBM is a tree ensemble method built on gradient boosting decision trees. It grows leaf-wise trees by selecting leaves with the largest decrease in loss and implements optimized histogram-based decision trees. Thus, it is exceedingly efficient and effective and was found to perform better than eXtreme Gradient Boosting (XGBoost) in such application areas as predicting insurance claims and flight delay, and ranking web search queries \cite{ke2017lightgbm}. Second, we used SHAP (a method that has good mathematical properties, such as consistency, missingness, and local accuracy \cite{lundberg2020local}) to identify the most important predictor variables (i.e., feature selection) to improve the performance of the LightGBM model, and to explain the effects of these factors on SA by calculating the contributions of the predictor variables in the LightGBM model using Shapley values from cooperative game theory \cite{shapley1953contributions}. 
In summary, the contributions of this paper are described as follows:
\begin{itemize}
  \item We built an explainable machine learning model with LightGBM and SHAP to predict driver SA in conditionally automated driving using eye-tracking data with reasonably high accuracy.
  \item We identified the most important eye-tracking measures in predicting SA in conditionally automated driving. 
  \item Our proposed method demonstrated the potential to measure and monitor SA in real time in conditionally automated driving and possibly in other dynamic environments. 
\end{itemize}

\section{RELATED WORK}
SA plays a critical role in the decision-making process across a wide range of applications. In conditionally automated driving, the vehicle might reach its operational limit during adverse driving conditions, in which it would request the driver to take over the driving task within a specific time budget \cite{ayoub2019manual}. It is of vital importance for the driver to maintain a good level of SA or resume his/her SA promptly in order to safely negotiate the driving scenarios \cite{zhou2019takeover,du2020psychophysiological}.

Measures for SA can be categorized into two types: subjective and objective measures. The subjective measures include the SART tool \cite{taylor2017situational}, and participant situation awareness questionnaire (PSAQ) \cite{strater2001measures}, and so on. For example, Petersen et al. \cite{petersen2019situational} used SART to measure drivers' SA during automated driving and found that by providing verbal information about the driving environment, drivers' SA was enhanced and so were their trust in automated driving and NDRT performance. Karjanto et al. \cite{karjanto2018effect} used peripheral LED strips to provide information on the future action of the automated vehicle to enhance drivers' SA, as evidenced by SART measurements. Self-report tools, such as SART, are easy to administer. However, participants cannot self-report information that they are not aware of. 

Objective measures include performance and behavioral measures as well as process indices. SAGAT \cite{endsley1995measurement} measures one's knowledge about the task by means of freeze-probes. Although this is one of the most commonly used techniques in dynamic tasks (e.g., aviation), it has also received criticism as it interferes with the task \cite{de2019situation} and thus is often not the first choice for takeover studies in automated driving. For example, Köhn et al. \cite{kohn2019improving} considered SAGAT to measure SA, but decided against it due to its interruptions of the driving task, which could counteract the out-of-the-loop problem during the takeover process. Takeover performance measures, including takeover time and takeover quality, are also associated with SA. For example, a low level of SA was associated with a longer takeover time \cite{clark2017situational}, worse takeover decisions \cite{eriksson2018rolling}, and worse NDRT performance \cite{petersen2019situational} compared to a high level of SA. However, a limitation of performance measures is that they are only available after the takeover maneuver.

How participants process information when performing tasks can also be used to measure SA. Examples are the measurement of communication patterns, physiological responses, and visual patterns. Walch et al. \cite{walch2017car} suggested that cooperative interfaces should be designed in automated vehicles to provide human-machine bi-lateral communication to increase SA. Hirano et al. \cite{hirano2018effects} examined the effects of music and verbal communication (i.e., talking to passengers) on drivers' SA during partially automated driving and found no significant improvement in their SA. SA can also be assessed by participants' physiological responses, which are often linked to cognitive constructs, such as drowsiness and mental workload \cite{french2017psycho}. For example, Zhou et al. \cite{zhou2021predicting,zhou2020driver} used physiological measures (e.g., heart rate, heart rate variability, and respiration rate) to detect participants' drowsiness and drowsiness transitions in highly automated driving to indicate their SA. French et al. \cite{french2017psycho} applied EEG to measure the three levels of SA. Zhang et al. \cite{zhang2020physiological} found that EEG was sensitive to changes in SA, and they also summarized the associations between other physiological measures and SA by reviewing previous studies.

Eye movement patterns enabled by eye trackers are also widely used to indicate SA.  In driving safety research, many crashes happen because drivers fail to look at the right objects at the right time or fail to project what the next move will be \cite{deng2016does}. In a study in which participants had to watch six dials with moving pointers, De Winter et al. \cite{de2019situation} found that visual-sampling scores (defined as the percentage of pointer threshold crossings for which participants fixated on the dial, within a threshold of 0.5 $s$ of the threshold crossing) correlated better with task performance than SAGAT. Yang et al. \cite{yang2017method} found that participants' visual scanning patterns, particularly eyes-on-road time had a marginally significant effect on takeover quality, i.e., the more one glanced on the road, the better one's SA tended to be. 
Braunagel et al. \cite{braunagel2017ready} found that eyes-on-road gazes alone could be used to predict takeover readiness with about 60\% accuracy. Lu et al. \cite{lu2017much} used eye-tracking measures, including net dwell time proportion in four different areas of interest and glance frequency on three mirrors (one rear-view and two side mirrors) to gain a deeper understanding of how participants resumed SA as a function of time during the takeover process. One of the merits of eye-tracking data is that it can be unobtrusively collected in real time, which provides the potential to monitor drivers' SA continuously in automated driving. Although driver state monitoring in driving has been widely studied using eye-tracking and other data \cite{dong2010driver}, few researchers have attempted to monitor and predict SA in automated driving using eye-tracking data.  

\begin{table*}[]
\renewcommand{\arraystretch}{1.3}
\caption{Predictor variables used in predicting situation awareness.}
\label{tab:predictor}
\begin{tabularx}{\textwidth}{lll}
    \hline
    \hline
Measures & Unit & Explanation \\
    \hline
1. age &
  year &
  age of the participants \\
2. gender &
  - &
  male = 1; female = 0 \\
3. yearDriving &
  year &
  years of driving experience \\
4. drivingFrequency &
  - &
  \makecell[l]{driving frequency in last 12 months: 
  every day driving = 1; 4-6 days/week= 2; 1-3 days/week = 3;\\ 
  once a month to once a week = 4; less than once/month = 5; never = 6} \\
5. videoLength &
  s &
  length of the video \\
6. decisionTime &
  s &
  time needed to make a decision \\
7. decisionMade &
  - &
  \makecell[l]{decision supposed to be made for each
  driving scenario: no need to take over = 1; evade left = 2; \\evade right = 3; 
  brake = 4} \\
8. correctDecision &
  - &
  \makecell[l]{correct decision made for each driving 
  scenario: no decision = 0; no need to 
  take over = 1; evade left = 2; \\evade 
  right = 3; evade left or right = 4; 
  brake only = 5} \\
9. danger &
  \makecell[l]{Likert
  scale} &
  \makecell[l]{Likert scale on "The situation 
  was dangerous": from completely 
  disagree = 0 to completely agree = 100} \\
10. difficulty &
    \makecell[l]{Likert
  scale} &
  \makecell[l]{Likert scale on "The rebuilding 
  task was difficult": from completely 
  disagree = 0 to completely agree = 100} \\
11. carPlacedLeft &
  - &
  \makecell[l]{number of cars placed on the left of the ego-vehicle} \\
12. carPlacedRight &
  - &
  \makecell[l]{number of cars placed on the right of 
  the ego-vehicle} \\
13. numS &
  - &
  \makecell[l]{number of the saccades of the 
  participant in one video} \\
14. sAmpMean &
  pixels &
  \makecell[l]{average saccade amplitude of the 
  participant in one video} \\
15. sAmpStd &
  pixels &
  \makecell[l]{standard deviation of saccade 
  amplitudes of the participant in one video} \\
16. sAmpMax &
  pixels &
  \makecell[l]{maximum value of saccade 
  amplitudes of the participant in one video} \\
17. numF &
  - &
  \makecell[l]{number of fixations of the 
  participant in one video} \\
18. fMean &
  ms &
  \makecell[l]{average fixation duration of the 
  participant in one video} \\
19. fStd &
  ms &
  \makecell[l]{standard deviation of fixation 
  duration of the participant in one video} \\
20. fMax &
  ms &
  \makecell[l]{maximum value of fixation duration
  of the participant in one video} \\
21. backMirror &
  - &
  \makecell[l]{number of fixations of the participant
  on the rear-view mirror in one video} \\
22. leftMirror &
  - &
  \makecell[l]{number of fixations of the participant
  on the left mirror in one video} \\
23. rightMirror &
  - &
  \makecell[l]{number of fixations of the participant
  on the right mirror in one video} \\
24. road &
  - &
  \makecell[l]{number of fixations of the participant
  on the road in one video} \\
25. sky &
  - &
  \makecell[l]{number of fixations of the participant
  on the sky in one video} \\
26. pupilChange &
  mm &
  \makecell[l]{pupil size change between the end of
  and the beginning of the video} \\
27. pupilMean &
  mm &
  \makecell[l]{average value of the pupil diameter
  of the participant in one video} \\
28. pupilStd &
  mm &
  \makecell[l]{standard deviation of the pupil 
  diameter of the participant in one video}\\
    \hline
    \hline
\end{tabularx}
\end{table*}

\section{DATASET}
We used the dataset collected in \cite{lu2020take} with 32 participants (29 males, 3 females) between 22 and 29 years old ($M = 24.2, SD = 1.8$). Each participant viewed 33 videos created using Prescan 8.0.0 (TASS International, The Netherlands) at 1080p with a frame rate of 20 $Hz$. The lengths of the videos ranged from 1  to 20 $s$ (1, 3, 6, 9, 12, or 20 $s$) featuring a conditionally automated vehicle from a driver's perspective in a three-lane driving scenario with a total number of 5 or 6 vehicles. All vehicles were randomly selected from 13 colors and 10 vehicle models. There were one to two vehicles in each lane and two to four vehicles in front of the ego-vehicle. Each vehicle was driving at one of three constant speeds, i.e., 80, 100, or 120 $km/h$, and there were zero to three vehicles driving at 80 $km/h$, one to three vehicles driving at 100 $km/h$, and one to three vehicles driving at 120 $km/h$. The ego-vehicle was always driving 100 $km/h$. The farthest distance between the ego-vehicle and other vehicles was 50-80 $m$. No vehicles performed lane changing in the videos.

Of all the driving scenarios, 16 required the participant to take over control from the automated vehicle due to a vehicle decelerating at 5 $m/s^2$ at the start of the video. Because of the dynamic constraints, such hazards would only be shown in videos with lengths of 1, 3, 6, and 9 $s$. The hazardous vehicle was standing still during the last second of the video. At the end of the video, the hazardous vehicle was 19-22 $m$ away from the ego-vehicle. After having watched the video, the participant was required to select the correct maneuver decision to avoid a collision. The decision options were ‘Evade left’, ‘Evade right’, ‘Brake only’, and ‘No need to take over’. 

In order to create the ground truth of SA of the participants, three performance measures in recreating the driving scenarios were used, namely 1) the absolute difference between the true number of vehicles and the placed number of vehicles in the driving scenario, 2) the error percentage of the distance between correctly placed vehicles and the true vehicles normalized to a scale from 0\% (perfect  placement) to 100\% (worst placement), and 3) total speed difference between correctly placed vehicles and the true vehicles, calculated by comparing the speed difference (equal, faster, or slower) between the ego-vehicle and others. Due to the different units involved in the performance measures, we normalized all the three error scores inversely into a scale from 0 to 1 and placed equal weights to create a global SA score between 0 (worst SA) and 1 (perfect SA). Because the ground truth of SA was a continuous variable normalized between 0 and 1, we modeled the SA prediction as a regression problem below.

To collect eye-tracking data of the participants, an EyeLink 1000 Plus (SR Research, Canada) eye tracker was used. It recorded participants' eye movements at a sampling rate of 2000 $Hz$, and was located 35 $cm$ in front of the 24-inch monitor. All the eye-tracking measures and other related predictor variables used as the input of the machine learning model are summarized in Table I. The first 12 measures are non-eye-tracking measures. In order to obtain the eye-tracking measures, we used the algorithms based on our previous study \cite{eisma2018visual}. We used a minimum fixation duration of 40 $ms$ to detect fixations (the minimum fixation observed was 93 $ms$) and a minimum speed threshold of 2000 $pixels/s$, a minimum duration of 15 $ms$, and a maximum duration of 150 $ms$ to detect saccades. In order to calculate pupil diameter, pupil areas were first preprocessed with a moving mean filter with a window size of 100 samples, and then blinks were identified using a threshold of 200 $ms$. The pupil areas were interpolated linearly for the blinks, and then a median filter was applied before pupil areas were converted to pupil diameter. We selected these eye-tracking measures as predictor variables mostly as a data-driven approach with the help of studies mentioned above. We did not include fixation duration measures for individual areas of interest because they were highly correlated with the number of fixations in the respective areas of interest, except on the road, which was highly correlated with the overall mean fixation duration. In this research, we will compare two types of prediction models, i.e., 1) the model that used all the measures and 2) the model that only used eye-tracking measures.

\section{Predicting SA with LightGBM and SHAP}
\subsection{LightGBM}
We used LightGBM \cite{ke2017lightgbm} to predict SA as a regression problem. LightGBM is an exceedingly efficient and effective ensemble machine learning model built on gradient boosting decision trees. It predicted SA by adding a large number of decision trees sequentially as follows: 

\begin{equation} \label{GBDT}
    \begin{split}
    \hat{y}_i^{0} &= 0\\
    \hat{y}_i^{1} &= f_1(\mathbf{x}_i)=\hat{y}_i^{0}+f_1(\mathbf{x}_i)\\
    ...\\
    \hat{y}_i^{j} &= \sum_{k=1}^j f_k(\mathbf{x}_i)=\hat{y}_i^{j-1}+f_j(\mathbf{x}_i),
    \end{split}
\end{equation}
where $\mathbf{x}_i = [x_{i1},x_{i2},...,x_{iP}], 1\leq i\leq N, N = 32\times 33 = 1056$ is the $i$-th input vector of all the predictor variables, $P=28$ (or $16$) is the total number of the (eye-tracking related) predictor variables in Table I, $N$ is the total number of the samples in the dataset, $\hat{y}_i^{j}$ is the predicted SA for $\mathbf{x}_i$ at $j$-th iteration, and $f_j$ is the  $j$-th trained decision tree. The algorithm used this process to minimize the objective function as follows, i.e., 
\begin{equation} \label{GBDT_Obj}
L^{j}(\theta) = \sum_{i}^{n}l(y_i,\hat{y}_i^{j})+\sum_{j=1}^{N}\Omega(f_j),
\end{equation}
where $l(y_i,\hat{y}_i^{j})$ is the loss function, being a combination of mean squared error and mean absolute error. The regularization term was used to reduce overfitting of the model by controlling its complexity. LightGBM used two strategies to improve the efficiency and effectiveness of the original gradient boosting decision trees. First, in the training process, in order to increase the efficiency of splitting the input recursively based on the information gain of the input, LightGBM utilized a leaf-wise method based on the so-called gradient-based one-side sampling (GOSS) strategy. It only kept the input with the most contributions to the information gain with large gradients and abandoned the input with small gradients randomly. Second, LightGBM bundled features that were nearly exclusive to each other to improve the efficiency and effectiveness of the model. We set the hyperparameters of the model as follows, without fine-tuning in the training and testing process: 'boosting\_type': 'goss', 'objective': 'regression', 'metric': {'l2', 'l1'}, 'num\_leaves': 100, 'learning\_rate': 0.05, 'bagging\_freq': 5, 'early\_stopping\_rounds': 100, and 'num\_boost\_round': 5000.

\subsection{SHAP}
SHAP \cite{lundberg2017unified} explains a machine learning model with desirable mathematical foundation, namely 1) local accuracy, 2) missingness, and 3) consistency. It defines the explanation model $g(\mathbf{x'})$ as a linear addition of the input variables to interpret the original function $f(\mathbf{x})$ as follows: 
\begin{equation} \label{addition}
f(\mathbf{x})=g(\mathbf{x'})=\varphi_{0}+\sum_{p=1}^{P}\varphi_{p}x'_{p},
\end{equation}
where $P$ is the total number of input variables, $\varphi_{0}$ is the bias when all input variables are not existing, $\mathbf{x} = h_\mathbf{x}(\mathbf{x'})$ with the mapping function, $h_\mathbf{x}$, and $\varphi_{p} \in R$ is the contribution to the prediction of the $p$-th input variable. For local accuracy, whenever $ \mathbf{x} = h_\mathbf{x}(\mathbf{x'})$, the explanation model, $g(\mathbf{x'})$, equals to the original function, $f(\mathbf{x})$. For missingness, when the input variable is missing, it will have no impact on the model prediction, i.e., $x'_{p} = 0 \rightarrow \varphi_{p} = 0$, which is satisfied in Eq. (3). For consistency, it means that for any two functions, $f$ and $f'$, if $f'_{x}(\mathbf{z'})-f'_{x}(\mathbf{z'}\setminus p)\geq f_{x}(\mathbf{z'})-f_{x}(\mathbf{z'}\setminus p)$, then  $\varphi_{p}(f',\mathbf{x})\geq \varphi_{p}(f,\mathbf{x})$, where $f_x(\mathbf{z'})=f(h_x(\mathbf{z'})), \mathbf{z}\in \{0,1\}^P, \mathbf{z}\setminus p$ indicates $z_p = 0$. In order for this to hold, the only solution, based on the Shapley value \cite{shapley1953contributions} obtained from coalitional game theory, is 
\begin{equation} \label{main}
\varphi_{p}(f,\mathbf{x})=\sum_{\mathbf{z'}\subseteq \mathbf{x'}} \frac {|\mathbf{z'}|!\;(P-|\mathbf{z'}|-1)!}{P!}(f_{x}(\mathbf{z'})-f_{x}(\mathbf{z'}\setminus p)), 
\end{equation}
where $|\mathbf{z'}|$ is the number of non-zero variables in $\mathbf{z'}$, which is a subset of $\mathbf{x'}$, i.e., $\mathbf{z'}\subseteq \mathbf{x'}$. Eq. (4) calculates the Sharpley value, i.e., $\varphi_{p}$, which indicates its fair contribution to the prediction of the $p$-th input variable. According to \cite{lundberg2017unified}, the solution to Eq. (4) is known as SHAP values, i.e., $ f_{x}(\mathbf{z'}) = f(h_x(\mathbf{z'})) =  E[f(\mathbf{z})|\mathbf{z'}_S]$, where $\mathbf{z'}_S$ is the non-zero set of $\mathbf{z'}$. It thus provides unique additive feature importance measure and satisfies the three properties described above.

We used SHAP to 1) identify the importance of individual predictor variables (also used for feature selection in the LightGBM model) globally, 2) explain the main effects of important predictor variables on SA, and 3) explain individual prediction instances by identifying the contributions of individual predictor variable-value sets. However, it is computationally intensive to calculate SHAP values because of the exponential complexity in Eq. (4). For tree ensembles, such as LightGBM, a more efficient algorithm was proposed by Lundberg et al. \cite{lundberg2018consistent} with $O(TLD^2)$ time, where $T$ is the number of the trees, $L$ is the number of maximum leaves in any tree, and $D = logL$.

\section{Results}
\subsection{Prediction Results}
We used 10-fold cross-validation to examine the performance of LightGBM, which was compared with other regression models. We used three performance measures, including RMSE, MAE, and correlation coefficient between the predicted SA and the ground truth, defined as follows:
\begin{equation} \label{rmse}
   RMSE = \sqrt{\frac{\sum_{i=1}^N \big(y_{i}-\hat{y}_{i}\big)^2}{N}}, 
\end{equation}
\begin{equation} \label{mae}
   MAE = \frac{\sum_{i=1}^N |y_{i}-\hat{y}_{i}|}{N} , 
\end{equation}
\begin{equation} \label{Corr}
Corr. = \frac{\sum_{i=1}^{N} (\hat{y_i}-\bar{\hat{y}})(y_i -\bar{y})} {\sqrt{\sum_{i=1}^{N} (\hat{y_i}-\bar{\hat{y}})^2 \sum_{i=1}^{N}(y_i -\bar{y})^2}}, \\
\end{equation}
where \textit{N} is the total number of the samples in the dataset, $y_i$ is the $i$-th value of SA samples, $\mathit{\hat{y}_i}$ is the $i$-th predicted SA, $\mathit{\bar{y}}$ is the mean value of all the SA samples, and $\bar{\hat{y}}$ is the mean value of all the predicted SA results.

\begin{table}[]
\renewcommand{\arraystretch}{1.3}
\caption{Performance of the selected machine learning models with all the predictor variables.}
\label{tab:resultAll}
\centering
\begin{tabular}{llll}
\hline\hline
Model & RMSE & MAE & Corr. \\ \hline 
Linear Regression & 0.121 & 0.097 & 0.447 \\
Linear SVM & 0.122 & 0.097 & 0.436 \\
Quadratic SVM & 0.121 & 0.094 & 0.458 \\
Gaussian SVM & 0.115 & 0.089 & 0.529 \\
Medium Tree & 0.124 & 0.098 & 0.412 \\
Random Forest & 0.111 & 0.088 & 0.566 \\
XGBoost  & 0.110 & 0.087 & 0.788 \\
LightGBM (all) & \textbf{0.109} & \textbf{0.086} & \textbf{0.796} \\
LightGBM (best) & \textbf{0.108} & \textbf{0.085} & \textbf{0.801} \\
\hline
\hline 
\end{tabular}

Note that we normalized SA between 0 and 1 and used the Regression Learner App in Matlab 2020b to evaluate the performance of the models, except for XGBoost and LightGBM, which were conducted in Python 3.8 in JupyterLab 2.1.5. The same approach was also adopted to obtain the results in Table III. LightGBM (all) indicates all the predictor variables were included in the model and LightGBM (best) indicates only the top 14 predictor variables were included (see Fig. \ref{fig:PerformanceAll}). 
\end{table}

\begin{table}[]
\renewcommand{\arraystretch}{1.3}
\caption{Performance of the selected machine learning models with only eye-tracking related predictor variables.}
\label{tab:resultEye}
\centering
\begin{tabular}{llll}
\hline\hline
Model & RMSE & MAE & Corr. \\ \hline 
Linear Regression & 0.126 & 0.102 & 0.361 \\
Linear SVM & 0.127 & 0.102 & 0.346 \\
Quadratic SVM & 0.143 & 0.103 & 0.332 \\
Gaussian SVM & 0.124 & 0.097 & 0.400 \\
Medium Tree & 0.136 & 0.108 & 0.300 \\
Random Forest & 0.123 & 0.097 & 0.412 \\
XGBoost  & 0.127 & 0.100 & 0.691 \\
LightGBM (all) & \textbf{0.122} & \textbf{0.096} & \textbf{0.716} \\
LightGBM (best) & \textbf{0.121} & \textbf{0.096} & \textbf{0.719} \\
\hline
\hline 
\end{tabular}
\\LightGBM (all) indicates all the eye-tracking predictor variables were included in the model and LightGBM (best) indicates only the top 9  eye-tracking predictor variables were included (see Fig. \ref{fig:PerformanceEye}). 
\end{table}

\begin{figure} [bt!]
	\centering
	\subfloat[\label{fig:PerformanceAll}]{\includegraphics[width=1\linewidth]{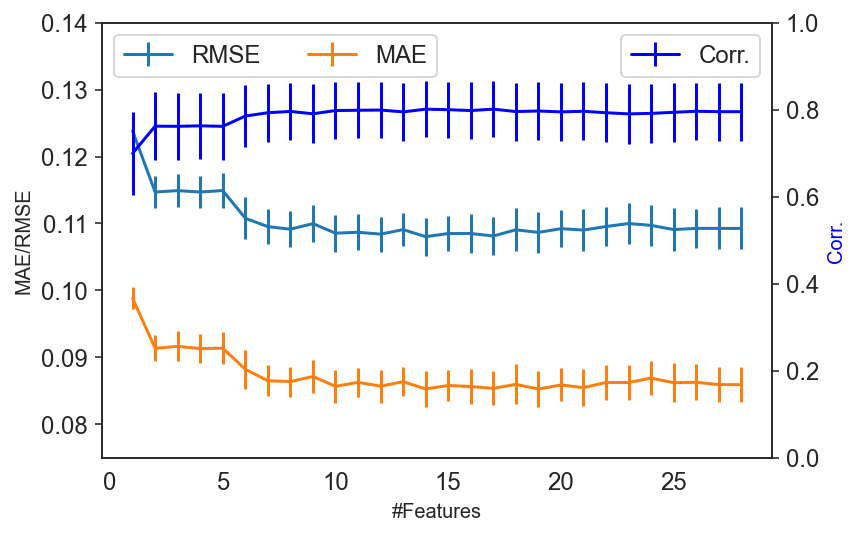}}
	\vspace{0pt}
	\subfloat[\label{fig:PerformanceEye}]{\includegraphics[width=1\linewidth]{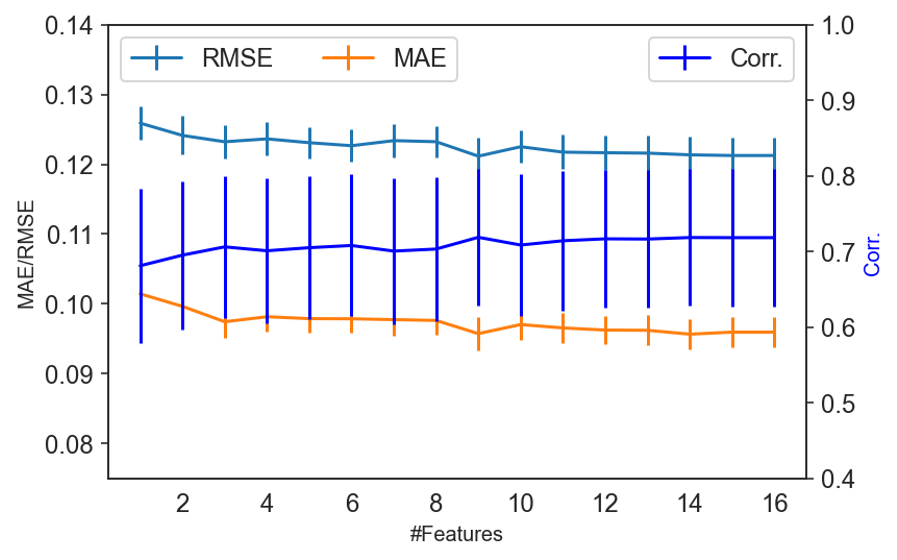}}
    \caption{Performance of the LightGBM models when predictor variables were sequentially selected from the most important one to the less important ones:  (a)  All predictor variables;  (b) Only eye-tracking related predictor variables. Note that the error bar was the standard error obtained in the ten-fold cross-validation process.}\label{fig:Performance}
\end{figure}

\begin{figure}
	\centering
	\subfloat[\label{fig:ImportanceAll}]{\includegraphics[width=1\linewidth]{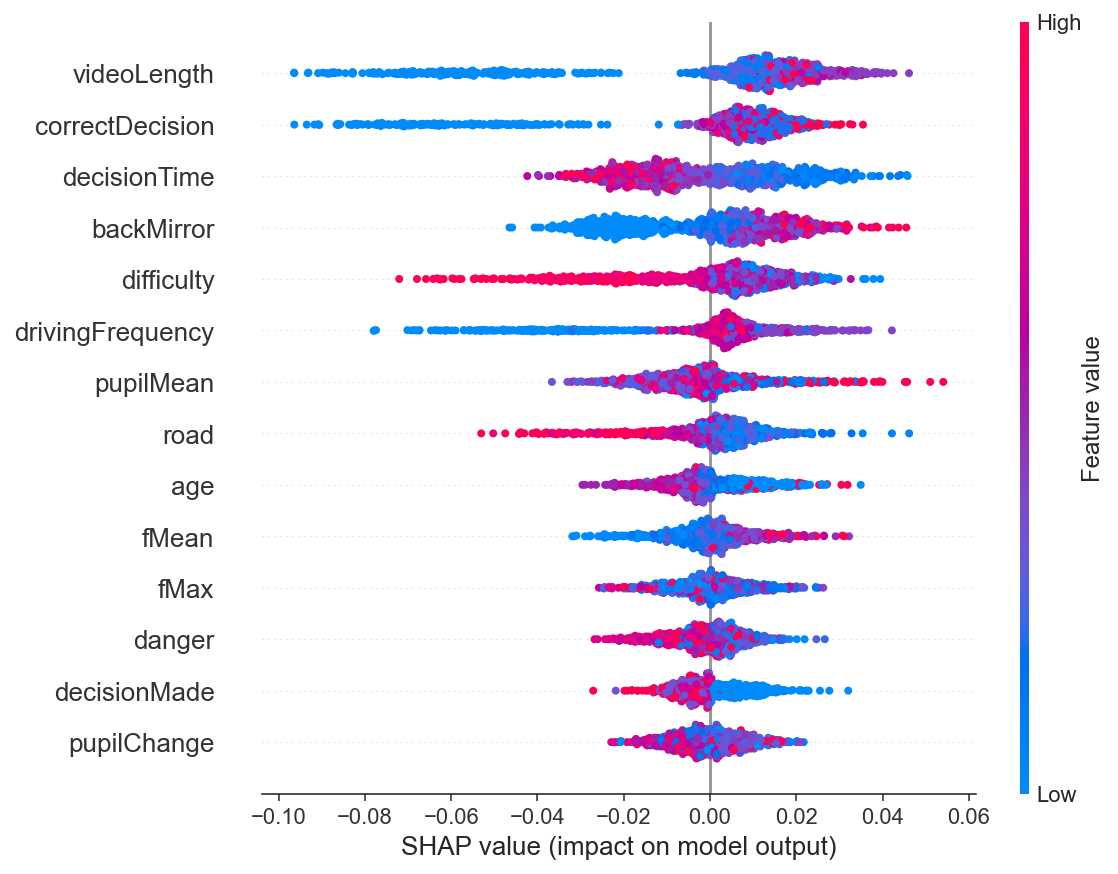}}
	\vspace{0pt}
	\subfloat[\label{fig:ImportanceEye}]{\includegraphics[width=1\linewidth]{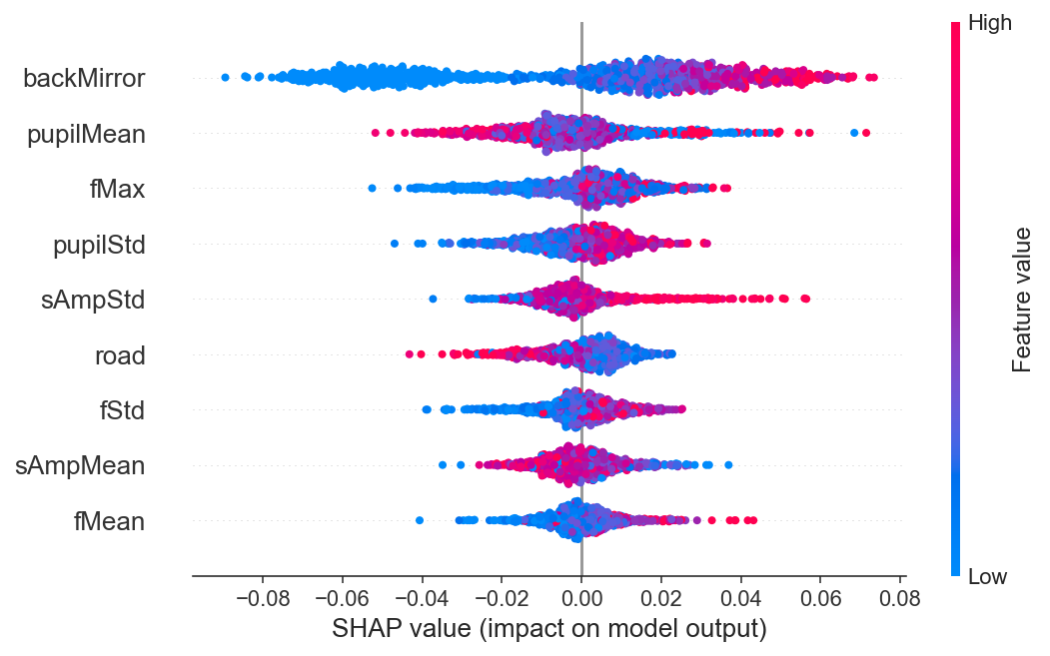}}
    \caption{Importance ranking of the predictor variables in the LightGBM model (best) produced by SHAP as shown in Table II and Table III: (a) When all the predictor variables were included and (b) when only eye-tracking related variables were included.}
\label{fig:importance}
\end{figure}

Table II shows the comparative performance of the selected regression models with all the 28 predictor variables in Table I and across the three performance measures (RMSE, MAE, and correlation coefficient), we found that LightGBM had the best performance, where LightGBM (all) included all the predictor variables and LightGBM (best) included only the top 14 predictor variables selected by SHAP (see Fig. \ref{fig:PerformanceAll} and Fig. \ref{fig:ImportanceAll}). Table III shows the comparative performance of selected regression models with all the 16 eye-tracking related predictor variables in Table I and across the three performance measures (RMSE, MAE, and correlation coefficient), we found that LightGBM had the best performance, where LightGBM (all) included all the eye-tracking related predictor variables and LightGBM (best) included only the top 9 predictor variables selected by SHAP (see Fig. \ref{fig:PerformanceEye} and Fig. \ref{fig:ImportanceEye}). Note that in order to obtain the best performance, we sequentially selected one variable at a time from the most important one to the less important ones. Fig. \ref{fig:Performance} shows how the performance changes when more predictor variables were included in the LightGBM model. As shown in Fig. \ref{fig:Performance}, after the first several predictor variables were included in the model, the performance seemed stabilized. This might be because the variables included were from the most important one to the least important ones in terms of their contributions to predicting SA. As a result, in Fig. \ref{fig:PerformanceAll}, the top 14 predictor variables (see Fig. \ref{fig:ImportanceAll}) were selected when the model had the best performance as shown in Table II; in Fig. \ref{fig:PerformanceEye}, the top 9 (see Fig. \ref{fig:ImportanceEye}) predictor variables were selected when the model had the best performance as shown in Table III. 


\begin{figure*} [hbt!]
\centering
\includegraphics[width=.97\linewidth]{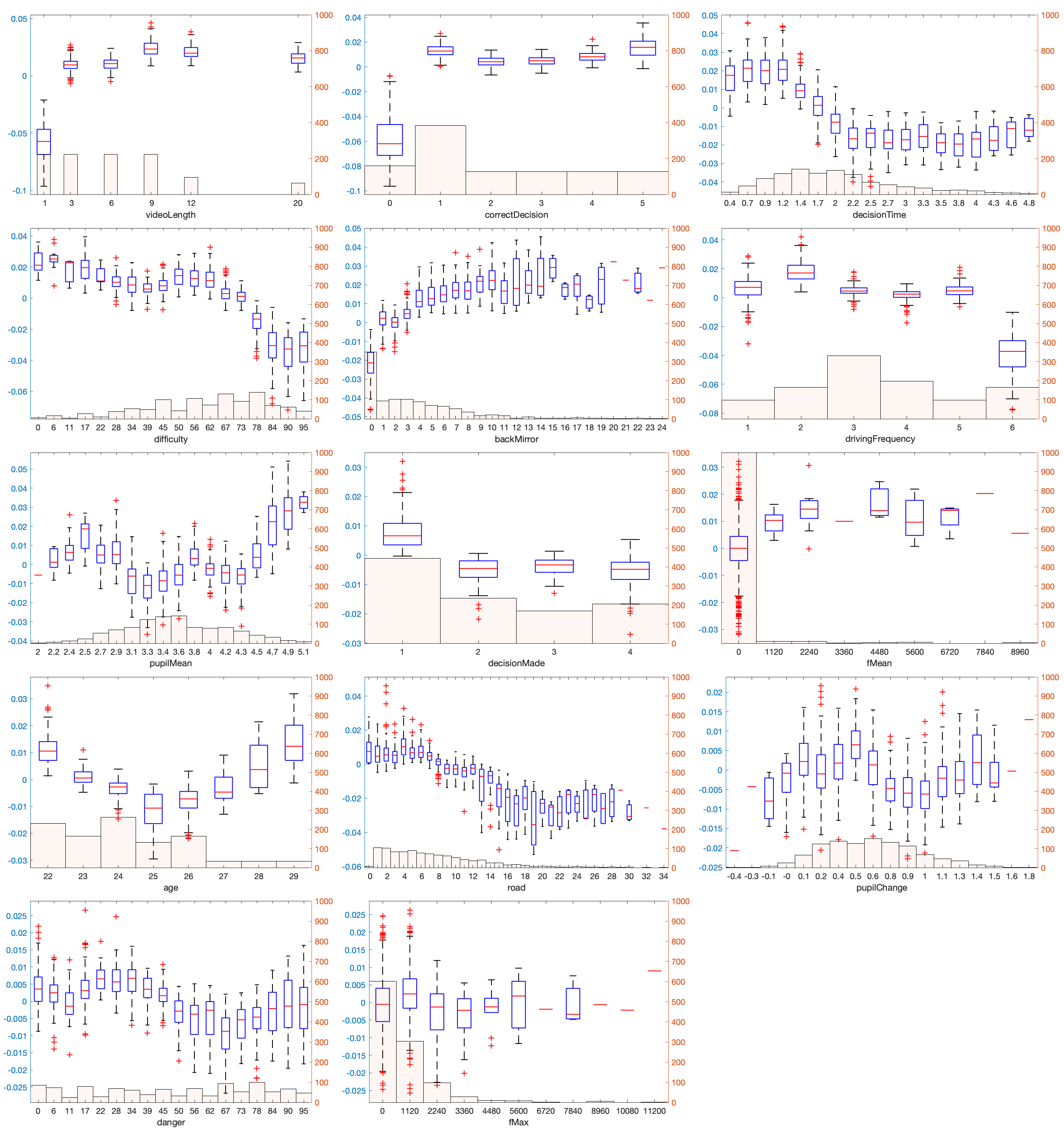}
\caption{The main effects of the top 14 most important predictor variables on SA. Note that the x-axis of the figure shows the values of the predictor variables, and the figure has two y-axes. The left y-axis indicates the SHAP values (i.e., the contributions to predicting SA). The box plots corresponding to the x-axis and the left y-axis indicate the main effects of the predictor variables on predicting SA. The right y-axis indicates the number of samples. The histogram corresponding to the x-axis, and the right y-axis shows the distributions of the dataset. This also applies to Fig. \ref{fig:mainEffectEye} below.}
\label{fig:mainEffectAll}
\end{figure*}

\subsection{SHAP Explanation}
\subsubsection{Feature Importance}
We computed the SHAP values in the same unit space as SA during the training and testing (i.e., 10-fold cross-validation) process of LightGBM. For each fold, we used 10\% of the test data to calculate the SHAP values, and this process was repeated 10 times so that each sample (i.e., $\mathbf{x}_i$ in Eq.(1)) in the dataset was calculated. Fig. \ref{fig:importance} shows the importance of the predictor variables with their global impact, $\sum_{i=1}^N|\varphi_{ip}|$, (i.e., the sum of the absolute SHAP values of all the instances of the $p$-th variable) on the LightGBM (best) prediction model, where Fig. \ref{fig:ImportanceAll} shows the results when all the predictor variables were included in the LightGBM model in Table II and Fig. \ref{fig:ImportanceEye} shows the results when only eye-tracking related variables were included in the LightGBM model in Table III. In each figure, each dot indicates one SHAP value, i.e., $\varphi_{ip}$ for one specific variable, and the figure has four aspects that can help interpret it: 1) The colors changing from blue (low) to red (high) indicate the value of the variable changing from low to high; 2) the horizontal axis shows the SHAP values with regard to the baseline value, indicating the effects (positive values increase SA, while negative values decrease SA) of the predictor variables on SA; 3) the vertical axis sorts the importance of the predictor variables (those at the top are more important than those below. For example, the most important is videoLength followed by correctDecision in Fig. \ref{fig:ImportanceAll}) by their global impacts on SA, i.e., $\sum_{i=1}^N|\varphi_{ip}|$; and 4) the shape of the horizontal violin plot of each predictor variable shows the distributions of the samples of that variable in the dataset. 
\begin{figure*} [hbt!]
\centering
\includegraphics[width=.97\linewidth]{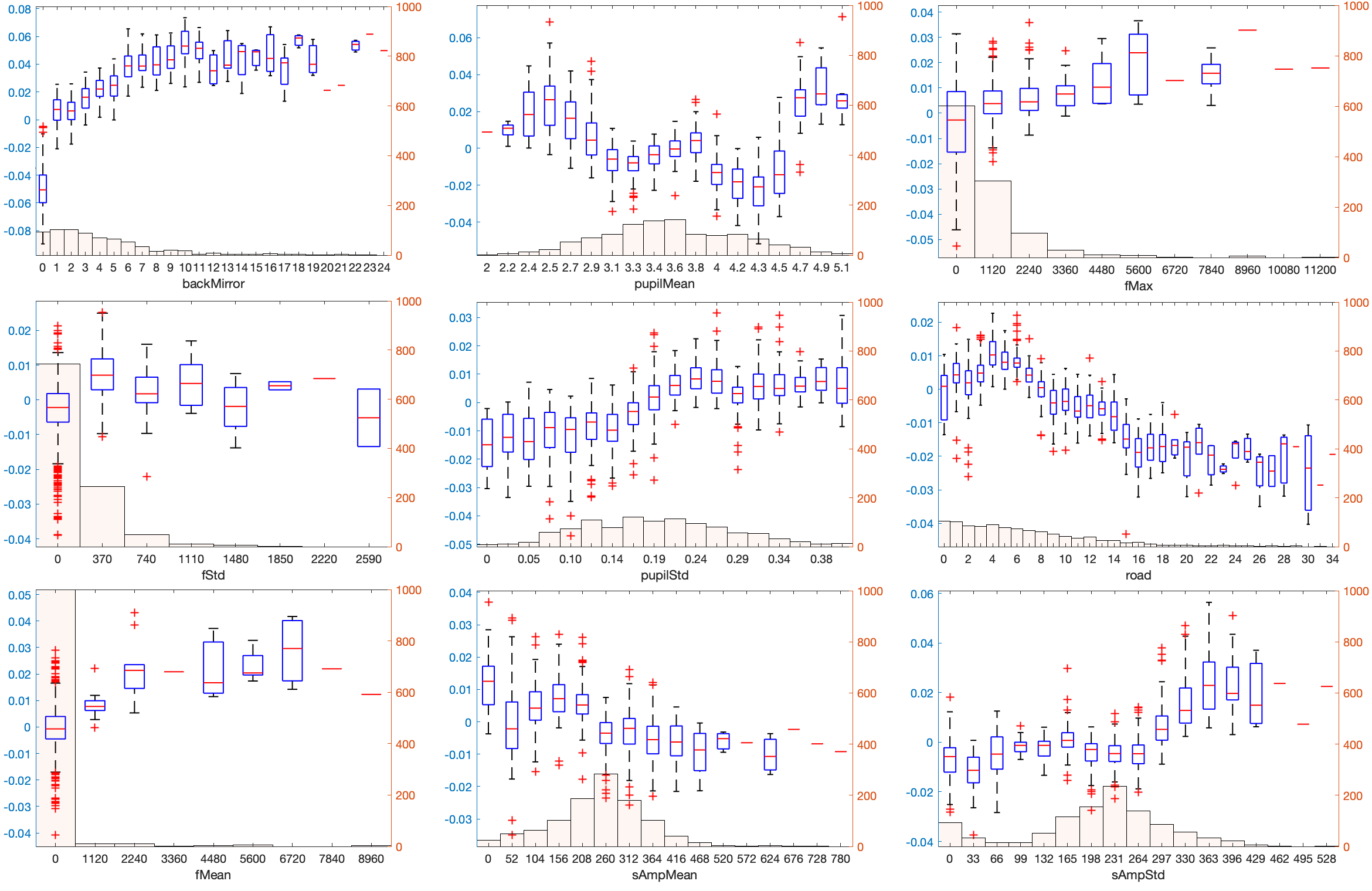}
\caption{The main effects of top 9 most important predictor variables related to eye-tracking measures on SA. Note that box plots correspond to the left y axis in SHAP values, indicating the main effects (positive ones increased SA while negative ones decreased SA)  while the histograms correspond to the right y axis, indicating the distribution of the samples of each predictor variable.}
\label{fig:mainEffectEye}
\end{figure*}

\subsubsection{Main effects of predictor variables on SA}
In order to understand the effects of important predictor variables on SA, we showed the main effect plots produced by SHAP in Figs. \ref{fig:mainEffectAll} and \ref{fig:mainEffectEye}. We plotted them in a series of box plots (the left y axis) to show the variations of SHAP values and histograms (the right y axis) to show sample distributions of the variables in the dataset. In order to create box plots for continuous variables, we grouped their values into an appropriate number of bins based on their distributions. For example, for the variable decisionTime, we discretized it into 18 bins, consistent with its histogram. Note that  Fig. \ref{fig:mainEffectAll} was produced using the LightGBM (best) model in Table II and Fig. \ref{fig:mainEffectEye} was produced using the LightGBM (best) in Table III. We also calculated Pearson's product-moment or Spearman's Rho rank-order correlation coefficients between the continuous or nominal predictor variables and their SHAP values, which indicate the individual contributions of variable-value sets to SA.

Fig. \ref{fig:mainEffectAll} shows the main effects of the selected top 14 predictor variables (see Fig. \ref{fig:ImportanceAll}). 
The most important predictor variable in Fig. \ref{fig:mainEffectAll} is videoLength ($r = .798, p = .000$), and five different lengths were included in the dataset. Compared to others, videos of 1 s reduced participants' SA by 0.02 to 0.09, while videos of other lengths increased participants' SA overall. However, there was not much difference in influencing SA among the videos whose lengths were longer than 3 s. The second most important variable is correctDecision ($r = .292, p = .000$). Compared to the trials in which participants did not make decisions ("0" in the figure), those in which participants made correct decisions had better SA. The third variable is decisionTime ($r = -.763, p = .000$), where a shorter decision time was associated with better SA. The fourth variable is difficulty ($r = -.746, p = .000$). The more difficult to rebuild the driving scenario, the lower the levels of SA the participants had. The fifth variable is the number of fixations on backMirror ($r = .745, p = .000$). The more fixation numbers on the rear-view mirror, the better SA was. The majority of the participants had 10 or fewer fixations. The sixth variable is drivingFrequency ($r = -.678, p = .000$) and, generally speaking, it was negatively correlated with SA. Those who did not have any driving experience ("6" in the figure) had worse SA than others. The seventh variable is pupilMean ($r = .158, p = .000$). There was no straightforward relationship between the average pupil size and SA, possibly due to individual differences, such as sensitivity to light from the videos \cite{lu2020take}. The eighth variable is decisionMade ($r = -.768, p = .000$). Those who decided to take over from automated driving ("1" in the figure) increased their SA more than others. The ninth variable is fMean ($r = .286, p = .000$). Although there was an overall trend that the longer the fixation duration, the better the SA, one should be cautious because the majority of the fixation duration was below 1120 ms. The tenth variable is age ($r = -.376, p = .000$), which showed a V-shape relationship between age and SA. However, one should be cautious about interpreting it due to the small range of ages and the uneven distribution in the dataset. The eleventh variable is the number of fixations on the road ($r = -.771, p = .000$). The more one fixated on the road, the lower one's SA was. This might be because the participants needed to examine the mirrors to obtain an overall understanding of the driving scenarios and the more one fixated on the road, the less one fixated on the mirrors. The twelfth variable pupilChange ($r = -.226, p = .000$) appeared to have a similar pattern with pupilMean, with no straightforward relationships with SA. The thirteenth is danger ($r = -.450, p = .000$), which tended to be negatively correlated with SA. The last one is fMax ($r = .013, p = .677$), i.e., the maximum fixation duration. It looked similar to fMean to some degree, but the majority of the samples were more widely spread than fMean. However, this is the only variable that did not have a significant correlation with its SHAP values.

Fig. \ref{fig:mainEffectEye} shows the main effects of the selected top 9 predictor variables (see Fig. \ref{fig:ImportanceEye}). The first (backMirror, $r = .753, p = .000$), second (pupilMean, $r = -.232, p = .000$), third (fMax, $r = .378, p = .000$), sixth (road, $r = -.740, p = .000$), and seventh (fMean, $r = .410, p = .000$) variables are also shown in Fig. \ref{fig:mainEffectAll}. The model captured similar patterns between these predictor variables and SA, indicating its good consistency of SHAP in explaining LightGBM models. Other than these, the fourth is fStd ($r = .300, p = .000$), i.e., the standard deviation of fixation duration, which seemed to have a similar pattern with fMean and a similar distribution with fMax. The fifth variable is pupilStd ($r = .636, p = .000$), which seemed to be positively correlated with SA. The eighth variable is sAmpMean ($r = -.477, p = .000$), i.e., the average value of saccade amplitudes, which turned out to be negatively correlated with SA. The ninth variable is sAmpStd ($r = .518, p = .000$), i.e., the standard deviation of saccade amplitudes, which had a positive correlation with SA.

\begin{figure} [bt!]
	\centering
	\subfloat[\label{fig:ExampleAll}]{\includegraphics[width=1\linewidth]{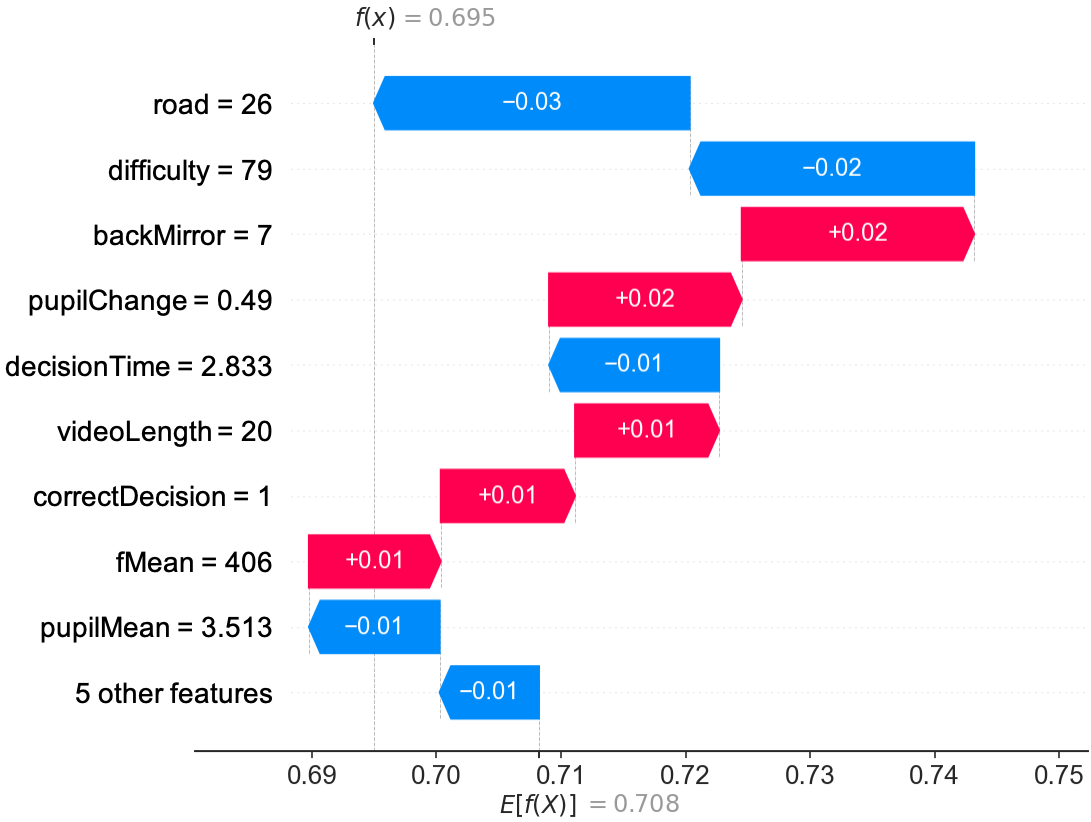}}
	\vspace{0pt}
	\subfloat[\label{fig:ExampleEye}]{\includegraphics[width=1\linewidth]{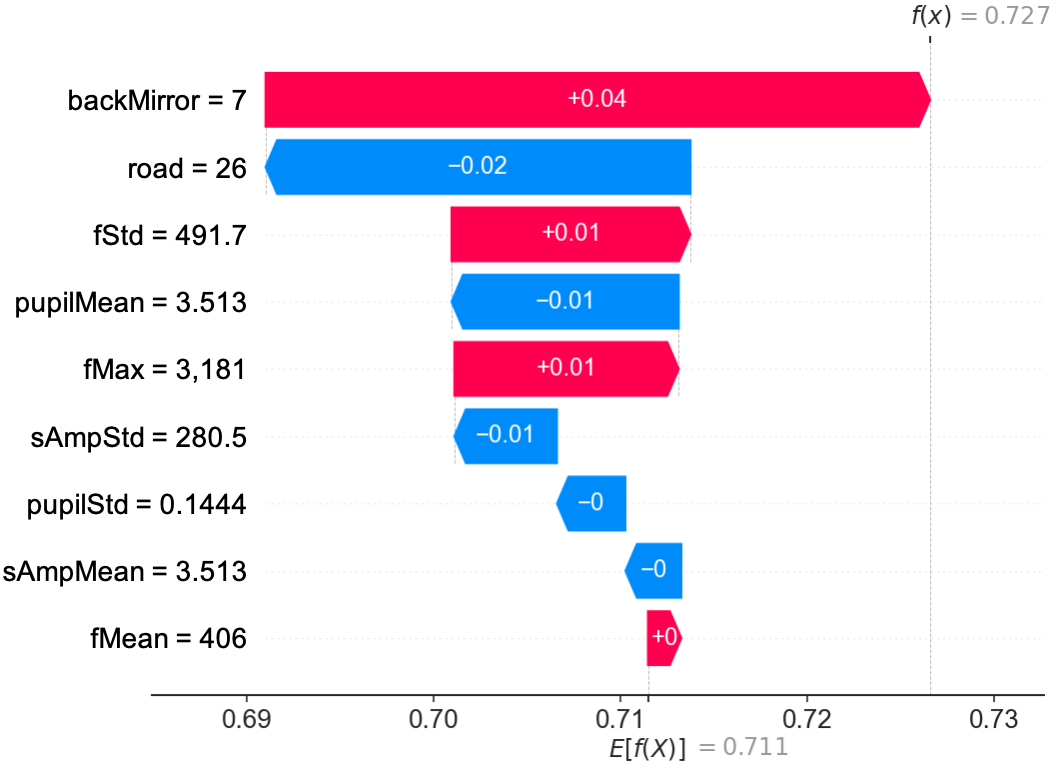}}
    \caption{Explaining individual instances using (a) LightGBM (best) model in Table II with the top 14 predictor variables and (b) LightGBM (best) model in Table III with the top 7 predictor variables related to eye-tracking data.}\label{fig:individual}
\end{figure}

\subsubsection{Explaining Individual Instances}
SHAP can also explain individual instances in the dataset to show the contributions of each variable-value set. For the two LightGBM (best) models in Table II and Table III, one specific example for each model is shown in Fig. \ref{fig:individual}.  In Fig. \ref{fig:ExampleAll}, the expected value $E[f(\mathbf{x})] = 0.708$ is the baseline SA value produced from the model and the dataset. The specific values in the individual instances increased (those in red) or decreased (those in blue) the predicted SA, through their own contributions. Those in blue, including "road = 26, difficulty = 79, decisionTime = 2.833, pupilMean = 3.513 and 5 other features", decreased the predicted SA, while those in red, including "backMirror = 7, pupilChange = 0.49, videoLength = 20, correctDecision = 1, fMean =  406", increased the predicted SA. The final predicted SA was 0.695, while the ground truth SA was 0.747. For the same instance, in Fig. \ref{fig:ExampleEye}, the baseline value is $E[f(\mathbf{x})] = 0.711$, which was slightly larger than that in the previous model. Those in blue, including "road = 26, pupilMean = 3.513, sAmpStd = 280.5, pupilStd = 0.1444, sAmpMean = 3.513", decreased the predicted SA while those in red, including "backMirror = 7, fStd = 491.7, fMean =  406.0", increased the predicted SA. The final predicted SA was 0.727, while the ground truth SA was 0.747. Note that the amounts increased or decreased by the same variable in these two models could be different. For example, in Fig. \ref{fig:ExampleAll} and Fig. \ref{fig:ExampleEye}, "backMirror = 7" increased the predicted SA by 0.02 and 0.04 while "road = 26" decreased the predicted SA by 0.03 and 0.02 with respect to the baseline SA value, respectively. These differences illustrate the effects and importance of the variables in each model. Variables listed at the top were more important than those listed below. For example, "backMirror = 7" was the most important variable when predicting SA with eye-tracking data alone in this instance in Fig. \ref{fig:ExampleEye}. However, the individual importance ranking in each instance can be different from the global importance ranking, as shown in Fig. \ref{fig:importance}.

\section{Discussions}
\subsection{Predicting SA}
In this study, we aimed to predict SA using eye-tracking data and other data using LightGBM and SHAP in the takeover process in conditionally automated driving. First, SA was measured by recreating the driving scenario during the takeover process. This approach was validated in \cite{lu2020take} through performance measures concerning car placement, distance, and speed to indicate participants' SA levels. This measurement of SA is similar to the SAGAT technique \cite{endsley1995measurement} in terms of recreating the driving scenarios in the virtual environment. However, one limitation of the current study is the ecological validity of the experiment since the data were collected in a low-fidelity setup (i.e., watching videos on a computer monitor).  

Second, we found that the LightGBM (best) model that used the top 14 predictor variables in Table II performed better than the LightGBM (best) model that used the top 9 eye-tracking related predictor variables in Table III. It is reasonable that when the model included predictor variables that directly related to SA \cite{lu2020take}, such as videoLength, correctDecision, decisionTime, and difficulty, it performed better as compared to when the model only had eye-tracking data as input. 
However, the major advantage of a model that only takes eye-tracking data as input is its potential to monitor and measure drivers' SA in real time without interfering with the task that the driver is performing. We used JupyterLab 2.1.5 with Python 3.8 for 10-fold cross-validation (including training and testing) for the optimal model (with nine eye-tracking measures, sample size = 1056), which took about 0.853 s on a MacBook Pro 13 (2.3 GHz Quad-Core Intel Core i7, 16 GB 3733 MHz LPDDR4X). Thus, predicting one sample took less than 1 ms. 

Third, we can possibly improve the performance of the model in the future when more data are available, such as behavioral (e.g., reaction time, eyes-on-road time) and physiological data (e.g., EEG), which were correlated with SA in previous studies \cite{miller2014situation,yang2017method,french2017psycho,zhang2020physiological}. Thus, during conditionally automated driving, drivers' SA might be monitored and measured in a non-intrusive way, and appropriate interventions may be provided when necessary in order to improve takeover performance. 

\subsection{Explaining SA Prediction}
In this study, we used SHAP to explain the LightGBM model by showing both the main effects of the most important predictor variables (see Figs. \ref{fig:mainEffectAll} and \ref{fig:mainEffectEye}) and individual explanations (see Fig. \ref{fig:individual}). First, for main effects, some of our results were consistent with those found in \cite{lu2020take}. For example, videos of 1 s resulted in impaired SA, decision accuracy was only weakly to moderately associated with SA, driving frequency was positively correlated with SA, and decision time was negatively correlated with SA. 

Second, more importantly, we built a prediction model with only eye-tracking data. We explored three types of eye-tracking measures: fixations, pupil diameter, and saccades. It seems that fixations were the most important in predicting SA, followed by pupil diameter and saccades. For example, both fixation numbers on the rear-view mirror (i.e., backMirror) and roads were selected in the model. Participants were found to view the rear-view mirror first to get an overall picture of the driving scenario \cite{lu2020take} and the more fixation numbers on it, the better their SA. Rensink \cite{rensink2000visual} showed that fixations were necessary to encode visual short-term memory by integrating sensory features into coherent object representations, which could be further transformed into long-term memory, especially when there were a sufficient number of fixations \cite{herten2017role}. Thus, when participants fixated more on relevant areas of interests, they remembered the driving scenarios better so that they had better SA. Moore and Gugerty \cite{moore2010development} also found that SA was positively associated with the number of fixations to relevant areas of interest, which is consistent with our finding regarding the number of fixations on the rear-view mirror. However, a larger number of fixations on the road decreased SA, which might be explained by the competitive relation between attention allocation between the rear-view mirror and the road. Moreover, an excessive number of fixations was found to be associated with difficulty in gathering information when the task demands and/or visual complexity were high \cite{jacob2003eye}. This finding might explain why SA decreased or leveled when the number of fixations on the road was greater than 6 and when the number of fixations on the rear-view mirror was greater than 11.

The general patterns of fixation duration measures (e.g., fMax, fStd, fMean) were similar in that the larger these values, the better SA. Fixation duration is known to be positively associated with the number of targets and dynamic memory load (within and above the limit of working memory capacity) \cite{meghanathan2015fixation}. Participants with a longer fixation duration were able to remember more vehicles (and also their location and speed). Furthermore, Lu et al. \cite{lu2017much} also indicated that participants were able to accurately estimate the total number of vehicles up to six in videos up to 20 s. Therefore, the vehicles to be remembered in order to recreate the driving scenarios did not exceed the capacity of the working memory of the participants.
This might explain the significant positive association between fixation duration and SA. However, the association was of moderate strength (correlation coefficients smaller than 0.5).  It was also observed that the distributions of these measures were similar to exponential distributions, which dramatically reduced the sample sizes as their values increased. Therefore, one should be cautious about interpreting the associations when the values are large. 

For the pupil diameter, pupilMean was more important than pupilStd, although pupilStd was found to be moderately correlated with its SHAP values. Meghanathan et al. \cite{meghanathan2015fixation} found that pupil diameter represented the number of targets only when it exceeded the capacity of working memory, and the fact that participants needed to identify up to six vehicles was not beyond the capacity of working memory \cite{lu2017much}. This, to some extent, might explain that pupilMean did not have a strong linear relationship with its SHAP values. In addition, pupil diameter is associated with other factors too, including individual differences and lighting fluctuations \cite{zhang2020physiological,lu2020take}. However, we did not find any studies reporting the association between pupilStd and SA. More research is needed to understand the role of pupil diameter in SA. 

The saccade measures had a similar pattern as the pupil measures (pupilMean vs. sAmpMean and pupilStd vs. sAmpStd). Saccades are indicative of attention shifts in visual search and scanning \cite{eckstein2017beyond}. Consistent with previous findings \cite{merchant2001evaluation}, saccade amplitudes were found to be negatively correlated with SA, indicating when the scan paths were shorter without searching across the driving scenario, it was more likely that participants identified the vehicle information more easily.  Our results indicate that when participants were scanning the driving scenario with larger saccade amplitude variations (i.e., high sAmpStd), they had a higher level of SA. This finding might be related to thorough visual search patterns to reduce errors. 

Third, we also examined the contributions of each variable-value sets to individual instance prediction in Fig. \ref{fig:individual}. Note the global importance of the predictor variables might not always be consistent with contributions to individual instances. We found the top three most important factors were "road = 26, difficulty = 29, and backMirror = 7" in Fig. \ref{fig:ExampleAll} and "backMirror =7, road = 26, and fStd = 491.7" in Fig. \ref{fig:ExampleEye}. Their effects on increasing and decreasing SA from the expected value of SA showed how the final prediction was reached. 

In summary, SHAP was able to successfully explain a tree-based ensemble machine learning model, LightGBM, and helped us identify the most important factors in predicting SA and their effects on SA. Such domain knowledge, as extracted from the black-box driver SA prediction models, can potentially help design human-machine systems to optimize the joint performance in conditionally automated driving. Furthermore, drivers can potentially calibrate their trust levels in automated vehicles and help accept and adopt them in the long run. 
  
\section{Conclusion and Future Work}
We aimed to predict SA using both LightGBM and SHAP during the takeover process in conditionally automated driving. By comparing with other selected machine learning models, LightGBM had the best performance by selecting the most important predictor variables identified by SHAP. The model that only took eye-tracking related predictor variables had reasonably good performance, with RMSE = 0.121, MAE = 0.096, and correlation coefficient = 0.719 when SA was aggregated using three performance measures (i.e., vehicle placement, distance, and speed estimation) and normalized between 0 and 1. Moreover, we identified the main effects of the selected predictor variables. Such domain knowledge can help us build real-time SA prediction models using non-intrusive eye-tracking measures. 

We acknowledge that the data collection was conducted in a low-fidelity setup. In the future, researchers should replicate the experiment in high-fidelity driving simulators or naturalistic driving to see if similar results can be obtained. More measures related to eye-tracking (e.g., blink rate, nearest neighbor index \cite{zhang2020physiological}) and other physiological and behavioral data can be included to see if performance can be improved. The participants involved in data collection were mainly engineering students with an unbalanced gender ratio. Future studies should recruit more participants with a large age range and a balanced gender ratio.


%




\bibliography{main}

\begin{thebibliography}{10}
\providecommand{\url}[1]{#1}
\csname url@samestyle\endcsname
\providecommand{\newblock}{\relax}
\providecommand{\bibinfo}[2]{#2}
\providecommand{\BIBentrySTDinterwordspacing}{\spaceskip=0pt\relax}
\providecommand{\BIBentryALTinterwordstretchfactor}{4}
\providecommand{\BIBentryALTinterwordspacing}{\spaceskip=\fontdimen2\font plus
\BIBentryALTinterwordstretchfactor\fontdimen3\font minus
  \fontdimen4\font\relax}
\providecommand{\BIBforeignlanguage}[2]{{%
\expandafter\ifx\csname l@#1\endcsname\relax
\typeout{** WARNING: IEEEtran.bst: No hyphenation pattern has been}%
\typeout{** loaded for the language `#1'. Using the pattern for}%
\typeout{** the default language instead.}%
\else
\language=\csname l@#1\endcsname
\fi
#2}}
\providecommand{\BIBdecl}{\relax}
\BIBdecl

\bibitem{sae2018taxonomy}
SAE, \emph{Taxonomy and Definitions for Terms Related to Driving Automation
  Systems for On-Road Motor Vehicles}.\hskip 1em plus 0.5em minus 0.4em\relax
  {SAE} International in United States, J3016--201806, Jun. 2018.

\bibitem{AYOUB2021102}
J.~Ayoub, X.~J. Yang, and F.~Zhou, ``Modeling dispositional and initial learned
  trust in automated vehicles with predictability and explainability,''
  \emph{Transportation Research Part F: Traffic Psychology and Behaviour},
  vol.~77, pp. 102 -- 116, 2021.

\bibitem{braunagel2015driver}
C.~Braunagel, E.~Kasneci, W.~Stolzmann, and W.~Rosenstiel, ``Driver-activity
  recognition in the context of conditionally autonomous driving,'' in
  \emph{2015 IEEE 18th International Conference on Intelligent Transportation
  Systems}.\hskip 1em plus 0.5em minus 0.4em\relax IEEE, 2015, pp. 1652--1657.

\bibitem{endsley1995toward}
M.~R. Endsley, ``Toward a theory of situation awareness in dynamic systems,''
  \emph{Human factors}, vol.~37, no.~1, pp. 32--64, 1995.

\bibitem{ayoub2019manual}
J.~Ayoub, F.~Zhou, S.~Bao, and X.~J. Yang, ``From manual driving to automated
  driving: A review of 10 years of {AutoUI},'' in \emph{Proceedings of the 11th
  International Conference on Automotive User Interfaces and Interactive
  Vehicular Applications}, 2019, pp. 70--90.

\bibitem{du2020examining}
N.~Du, F.~Zhou, E.~M. Pulver, D.~M. Tilbury, L.~P. Robert, A.~K. Pradhan, and
  X.~J. Yang, ``Examining the effects of emotional valence and arousal on
  takeover performance in conditionally automated driving,''
  \emph{Transportation Research Part C: Emerging Technologies}, vol. 112, pp.
  78--87, 2020.

\bibitem{zhou2019takeover}
F.~Zhou, X.~J. Yang, and X.~Zhang, ``Takeover transition in autonomous
  vehicles: A {Y}ou{T}ube study,'' \emph{International Journal of
  Human--Computer Interaction}, pp. 1--12, 2019.

\bibitem{du2020predicting}
N.~Du, F.~Zhou, E.~M. Pulver, D.~M. Tilbury, L.~P. Robert, A.~K. Pradhan, and
  X.~J. Yang, ``Predicting driver takeover performance in conditionally
  automated driving,'' \emph{Accident Analysis \& Prevention}, vol. 148, p.
  105748, 2020.

\bibitem{zeeb2016take}
K.~Zeeb, A.~Buchner, and M.~Schrauf, ``Is take-over time all that matters?
  {The} impact of visual-cognitive load on driver take-over quality after
  conditionally automated driving,'' \emph{Accident Analysis \& Prevention},
  vol.~92, pp. 230--239, 2016.

\bibitem{braunagel2017online}
C.~Braunagel, D.~Geisler, W.~Rosenstiel, and E.~Kasneci, ``Online recognition
  of driver-activity based on visual scanpath classification,'' \emph{IEEE
  Intelligent Transportation Systems Magazine}, vol.~9, no.~4, pp. 23--36,
  2017.

\bibitem{du2020predictingchi}
N.~Du, F.~Zhou, E.~Pulver, D.~Tilbury, L.~P. Robert, A.~K. Pradhan, and X.~J.
  Yang, ``Predicting takeover performance in conditionally automated driving,''
  in \emph{Extended Abstracts of the 2020 CHI Conference on Human Factors in
  Computing Systems}, 2020, pp. 1--8.

\bibitem{endsley1995measurement}
M.~R. Endsley, ``Measurement of situation awareness in dynamic systems,''
  \emph{Human Factors}, vol.~37, no.~1, pp. 65--84, 1995.

\bibitem{taylor2017situational}
R.~M. Taylor, ``Situational awareness rating technique (sart): The development
  of a tool for aircrew systems design,'' in \emph{Situational
  Awareness}.\hskip 1em plus 0.5em minus 0.4em\relax Routledge, 2017, pp.
  111--128.

\bibitem{sirkin2017toward}
D.~Sirkin, N.~Martelaro, M.~Johns, and W.~Ju, ``Toward measurement of situation
  awareness in autonomous vehicles,'' in \emph{Proceedings of the 2017 CHI
  Conference on Human Factors in Computing Systems}, 2017, pp. 405--415.

\bibitem{petersen2019situational}
L.~Petersen, L.~Robert, X.~J. Yang, and D.~Tilbury, ``Situational awareness,
  driver’s trust in automated driving systems and secondary task
  performance,'' \emph{SAE International Journal of Connected and Automated
  Vehicles}, vol.~2, no. 12-02-02-0009, 2019.

\bibitem{young2013missing}
K.~L. Young, P.~M. Salmon, and M.~Cornelissen, ``Missing links? the effects of
  distraction on driver situation awareness,'' \emph{Safety Science}, vol.~56,
  pp. 36--43, 2013.

\bibitem{molnar2017age}
L.~J. Molnar, ``Age-related differences in driver behavior associated with
  automated vehicles and the transfer of control between automated and manual
  control: {A} simulator evaluation,'' University of Michigan, Ann Arbor,
  Transportation Research Institute, Tech. Rep., 2017.

\bibitem{yoon2019non}
S.~H. Yoon and Y.~G. Ji, ``Non-driving-related tasks, workload, and takeover
  performance in highly automated driving contexts,'' \emph{Transportation
  Research Part F: Traffic Psychology and Behaviour}, vol.~60, pp. 620--631,
  2019.

\bibitem{de2019situation}
J.~C. de~Winter, Y.~B. Eisma, C.~Cabrall, P.~A. Hancock, and N.~A. Stanton,
  ``Situation awareness based on eye movements in relation to the task
  environment,'' \emph{Cognition, Technology \& Work}, vol.~21, no.~1, pp.
  99--111, 2019.

\bibitem{ke2017lightgbm}
G.~Ke, Q.~Meng, T.~Finley, T.~Wang, W.~Chen, W.~Ma, Q.~Ye, and T.-Y. Liu,
  ``Lightgbm: A highly efficient gradient boosting decision tree,'' in
  \emph{Advances in Neural Information Processing Systems}, 2017, pp.
  3146--3154.

\bibitem{lundberg2018explainable}
S.~M. Lundberg, B.~Nair, M.~S. Vavilala, M.~Horibe, M.~J. Eisses, T.~Adams,
  D.~E. Liston, D.~K.-W. Low, S.-F. Newman, J.~Kim \emph{et~al.}, ``Explainable
  machine-learning predictions for the prevention of hypoxaemia during
  surgery,'' \emph{Nature Biomedical Engineering}, vol.~2, no.~10, pp.
  749--760, 2018.

\bibitem{lundberg2020local}
S.~M. Lundberg, G.~Erion, H.~Chen, A.~DeGrave, J.~M. Prutkin, B.~Nair, R.~Katz,
  J.~Himmelfarb, N.~Bansal, and S.-I. Lee, ``From local explanations to global
  understanding with explainable ai for trees,'' \emph{Nature Machine
  Intelligence}, vol.~2, no.~1, pp. 2522--5839, 2020.

\bibitem{shapley1953contributions}
L.~S. Shapley, H.~Kuhn, and A.~Tucker, ``Contributions to the theory of
  games,'' \emph{Annals of Mathematics Studies}, vol.~28, no.~2, pp. 307--317,
  1953.

\bibitem{du2020psychophysiological}
N.~Du, X.~J. Yang, and F.~Zhou, ``Psychophysiological responses to takeover
  requests in conditionally automated driving,'' \emph{Accident Analysis \&
  Prevention}, vol. 148, p. 105804, 2020.

\bibitem{strater2001measures}
L.~D. Strater, M.~R. Endsley, R.~J. Pleban, and M.~D. Matthews, ``Measures of
  platoon leader situation awareness in virtual decision-making exercises,''
  TRW Inc. Fairfax VA Systems \& Information Technology Group, Tech. Rep.,
  2001.

\bibitem{karjanto2018effect}
J.~Karjanto, N.~M. Yusof, C.~Wang, J.~Terken, F.~Delbressine, and
  M.~Rauterberg, ``The effect of peripheral visual feedforward system in
  enhancing situation awareness and mitigating motion sickness in fully
  automated driving,'' \emph{Transportation Research Part F: Traffic Psychology
  and Behaviour}, vol.~58, pp. 678--692, 2018.

\bibitem{kohn2019improving}
T.~K{\"o}hn, M.~Gottlieb, M.~Schermann, and H.~Krcmar, ``Improving take-over
  quality in automated driving by interrupting non-driving tasks,'' in
  \emph{Proceedings of the 24th International Conference on Intelligent User
  Interfaces}, 2019, pp. 510--517.

\bibitem{clark2017situational}
H.~Clark, A.~C. McLaughlin, and J.~Feng, ``Situational awareness and time to
  takeover: exploring an alternative method to measure engagement with
  high-level automation,'' in \emph{Proceedings of the Human Factors and
  Ergonomics Society Annual Meeting}, vol.~61, no.~1.\hskip 1em plus 0.5em
  minus 0.4em\relax SAGE Publications Sage CA: Los Angeles, CA, 2017, pp.
  1452--1456.

\bibitem{eriksson2018rolling}
A.~Eriksson, S.~M. Petermeijer, M.~Zimmermann, J.~C. De~Winter, K.~J. Bengler,
  and N.~A. Stanton, ``Rolling out the red (and green) carpet: {Supporting}
  driver decision making in automation-to-manual transitions,'' \emph{IEEE
  Transactions on Human-Machine Systems}, vol.~49, no.~1, pp. 20--31, 2018.

\bibitem{walch2017car}
M.~Walch, K.~M{\"u}hl, J.~Kraus, T.~Stoll, M.~Baumann, and M.~Weber, ``From
  car-driver-handovers to cooperative interfaces: Visions for driver--vehicle
  interaction in automated driving,'' in \emph{Automotive User
  Interfaces}.\hskip 1em plus 0.5em minus 0.4em\relax Springer, 2017, pp.
  273--294.

\bibitem{hirano2018effects}
T.~Hirano, J.~Lee, and M.~Itoh, ``Effects of auditory stimuli and verbal
  communications on drivers' situation awareness in partially automated
  driving,'' in \emph{2018 57th Annual Conference of the Society of Instrument
  and Control Engineers of Japan (SICE)}.\hskip 1em plus 0.5em minus
  0.4em\relax IEEE, 2018, pp. 690--696.

\bibitem{french2017psycho}
H.~T. French, E.~Clarke, D.~Pomeroy, M.~Seymour, and C.~R. Clark,
  ``Psycho-physiological measures of situation awareness,'' in \emph{Decision
  Making in Complex Environments}.\hskip 1em plus 0.5em minus 0.4em\relax CRC
  Press, 2017, pp. 291--297.

\bibitem{zhou2021predicting}
F.~Zhou, A.~Alsaid, M.~Blommer, R.~Curry, R.~Swaminathan, D.~Kochhar,
  W.~Talamonti, and L.~Tijerina, ``Predicting driver fatigue in automated
  driving with explainability,'' \emph{arXiv preprint arXiv:2103.02162}, 2021.

\bibitem{zhou2020driver}
F.~Zhou, A.~Alsaid, M.~Blommer, R.~Curry, R.~Swaminathan, D.~Kochhar,
  W.~Talamonti, L.~Tijerina, and B.~Lei, ``Driver fatigue transition prediction
  in highly automated driving using physiological features,'' \emph{Expert
  Systems with Applications}, p. 113204, 2020.

\bibitem{zhang2020physiological}
T.~Zhang, J.~Yang, N.~Liang, B.~J. Pitts, K.~O. Prakah-Asante, R.~Curry, B.~S.
  Duerstock, J.~P. Wachs, and D.~Yu, ``Physiological measurements of situation
  awareness: a systematic review,'' \emph{Human Factors}, p. 0018720820969071,
  2020.

\bibitem{deng2016does}
T.~Deng, K.~Yang, Y.~Li, and H.~Yan, ``Where does the driver look?
  top-down-based saliency detection in a traffic driving environment,''
  \emph{IEEE Transactions on Intelligent Transportation Systems}, vol.~17,
  no.~7, pp. 2051--2062, 2016.

\bibitem{yang2017method}
Y.~Yang, M.~G{\"o}tze, A.~Laqua, G.~C. Dominioni, K.~Kawabe, and K.~Bengler,
  ``A method to improve driver’s situation awareness in automated driving,''
  \emph{Proceedings of the Human Factors and Ergonomics Society Europe}, pp.
  29--47, 2017.

\bibitem{braunagel2017ready}
C.~Braunagel, W.~Rosenstiel, and E.~Kasneci, ``Ready for take-over? a new
  driver assistance system for an automated classification of driver take-over
  readiness,'' \emph{IEEE Intelligent Transportation Systems Magazine}, vol.~9,
  no.~4, pp. 10--22, 2017.

\bibitem{lu2017much}
Z.~Lu, X.~Coster, and J.~De~Winter, ``How much time do drivers need to obtain
  situation awareness? a laboratory-based study of automated driving,''
  \emph{Applied Ergonomics}, vol.~60, pp. 293--304, 2017.

\bibitem{dong2010driver}
Y.~Dong, Z.~Hu, K.~Uchimura, and N.~Murayama, ``Driver inattention monitoring
  system for intelligent vehicles: A review,'' \emph{IEEE Transactions on
  Intelligent Transportation Systems}, vol.~12, no.~2, pp. 596--614, 2010.

\bibitem{lu2020take}
Z.~Lu, R.~Happee, and J.~C. de~Winter, ``Take over! a video-clip study
  measuring attention, situation awareness, and decision-making in the face of
  an impending hazard,'' \emph{Transportation Research Part F: Traffic
  Psychology and Behaviour}, vol.~72, pp. 211--225, 2020.

\bibitem{eisma2018visual}
Y.~B. Eisma, C.~D. Cabrall, and J.~C. de~Winter, ``Visual sampling processes
  revisited: replicating and extending senders (1983) using modern eye-tracking
  equipment,'' \emph{IEEE Transactions on Human-Machine Systems}, vol.~48,
  no.~5, pp. 526--540, 2018.

\bibitem{lundberg2017unified}
S.~M. Lundberg and S.-I. Lee, ``A unified approach to interpreting model
  predictions,'' in \emph{Advances in Neural Information Processing Systems},
  2017, pp. 4765--4774.

\bibitem{lundberg2018consistent}
S.~M. Lundberg, G.~G. Erion, and S.-I. Lee, ``Consistent individualized feature
  attribution for tree ensembles,'' \emph{arXiv preprint arXiv:1802.03888},
  2018.

\bibitem{miller2014situation}
D.~Miller, A.~Sun, and W.~Ju, ``Situation awareness with different levels of
  automation,'' in \emph{2014 IEEE International Conference on Systems, Man,
  and Cybernetics (SMC)}.\hskip 1em plus 0.5em minus 0.4em\relax IEEE, 2014,
  pp. 688--693.

\bibitem{rensink2000visual}
R.~A. Rensink, ``Visual search for change: A probe into the nature of
  attentional processing,'' \emph{Visual cognition}, vol.~7, no. 1-3, pp.
  345--376, 2000.

\bibitem{herten2017role}
N.~Herten, T.~Otto, and O.~T. Wolf, ``The role of eye fixation in memory
  enhancement under stress--an eye tracking study,'' \emph{Neurobiology of
  learning and memory}, vol. 140, pp. 134--144, 2017.

\bibitem{moore2010development}
K.~Moore and L.~Gugerty, ``Development of a novel measure of situation
  awareness: The case for eye movement analysis,'' in \emph{Proceedings of the
  Human Factors and Ergonomics Society Annual Meeting}, vol.~54, no.~19.\hskip
  1em plus 0.5em minus 0.4em\relax SAGE Publications Sage CA: Los Angeles, CA,
  2010, pp. 1650--1654.

\bibitem{jacob2003eye}
R.~J. Jacob and K.~S. Karn, ``Eye tracking in human-computer interaction and
  usability research: Ready to deliver the promises,'' in \emph{The Mind's
  Eye}.\hskip 1em plus 0.5em minus 0.4em\relax Elsevier, 2003, pp. 573--605.

\bibitem{meghanathan2015fixation}
R.~N. Meghanathan, C.~van Leeuwen, and A.~R. Nikolaev, ``Fixation duration
  surpasses pupil size as a measure of memory load in free viewing,''
  \emph{Frontiers in Human Neuroscience}, vol.~8, p. 1063, 2015.

\bibitem{eckstein2017beyond}
M.~K. Eckstein, B.~Guerra-Carrillo, A.~T.~M. Singley, and S.~A. Bunge, ``Beyond
  eye gaze: What else can eyetracking reveal about cognition and cognitive
  development?'' \emph{Developmental Cognitive Neuroscience}, vol.~25, pp.
  69--91, 2017.

\bibitem{merchant2001evaluation}
S.~Merchant, Y.~Kwon, T.~Schnell, T.~Etherington, and T.~Vogl, ``Evaluation of
  synthetic vision information system (svis) displays based on pilot
  performance,'' in \emph{20th DASC. 20th Digital Avionics Systems Conference
  (Cat. No. 01CH37219)}, vol.~1.\hskip 1em plus 0.5em minus 0.4em\relax IEEE,
  2001, pp. 2C1--1.

\end{thebibliography}



%


\vskip 0pt plus -1fil
\begin{IEEEbiography}[{\includegraphics[width=1in,height=1.55in,clip,keepaspectratio]{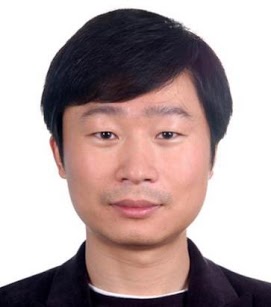}}]{Feng Zhou} received the Ph.D. degree in Human Factors Engineering from Nanyang Technological University, Singapore, in 2011 and the Ph.D. degree in Mechanical Engineering from Georgia Tech in 2014. He was a Research Scientist at MediaScience, Austin TX, from 2015 to 2017. He is currently an Assistant Professor with the Department of Industrial and Manufacturing Systems Engineering, University of Michigan, Dearborn. His main research interests include human factors, human-computer interaction, engineering design, and human-centered design.
\end{IEEEbiography}
\vskip 0pt plus -1fil

\begin{IEEEbiography}[{\includegraphics[width=1in,height=1.25in,clip,keepaspectratio]{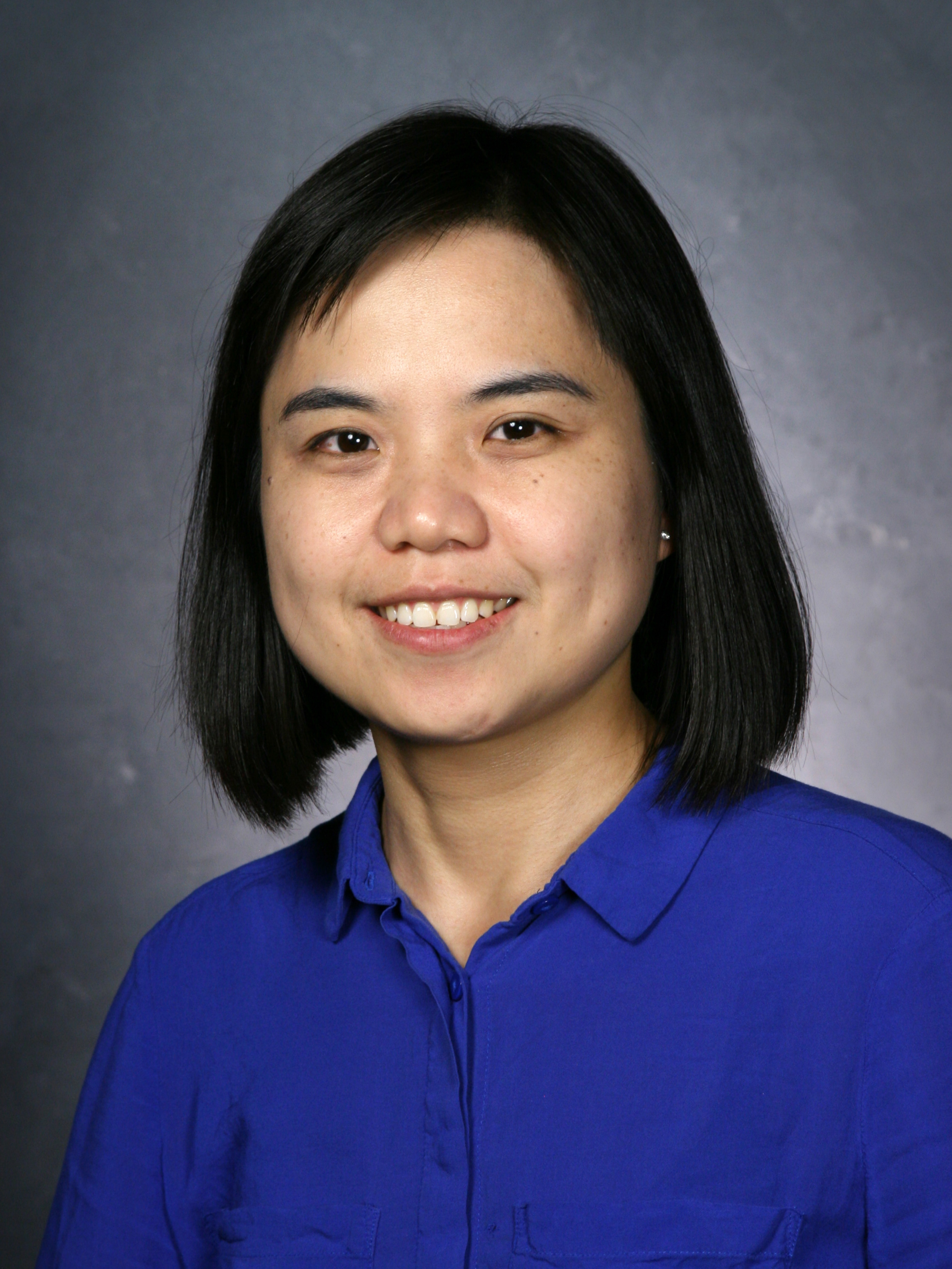}}]{X. Jessie Yang}
is an Assistant Professor at the Department of Industrial and Operations Engineering, University of Michigan Ann Arbor. She obtained her PhD in Mechanical and Aerospace Engineering (Human Factors) from Nanyang Technological University, Singapore in 2014. Dr. Yang’s research include human-autonomy interaction, human factors in high-risk industries and user experience design.
\end{IEEEbiography}
\vskip 0pt plus -1fil

\begin{IEEEbiography}[{\includegraphics[width=1in,height=1.25in,clip,keepaspectratio]{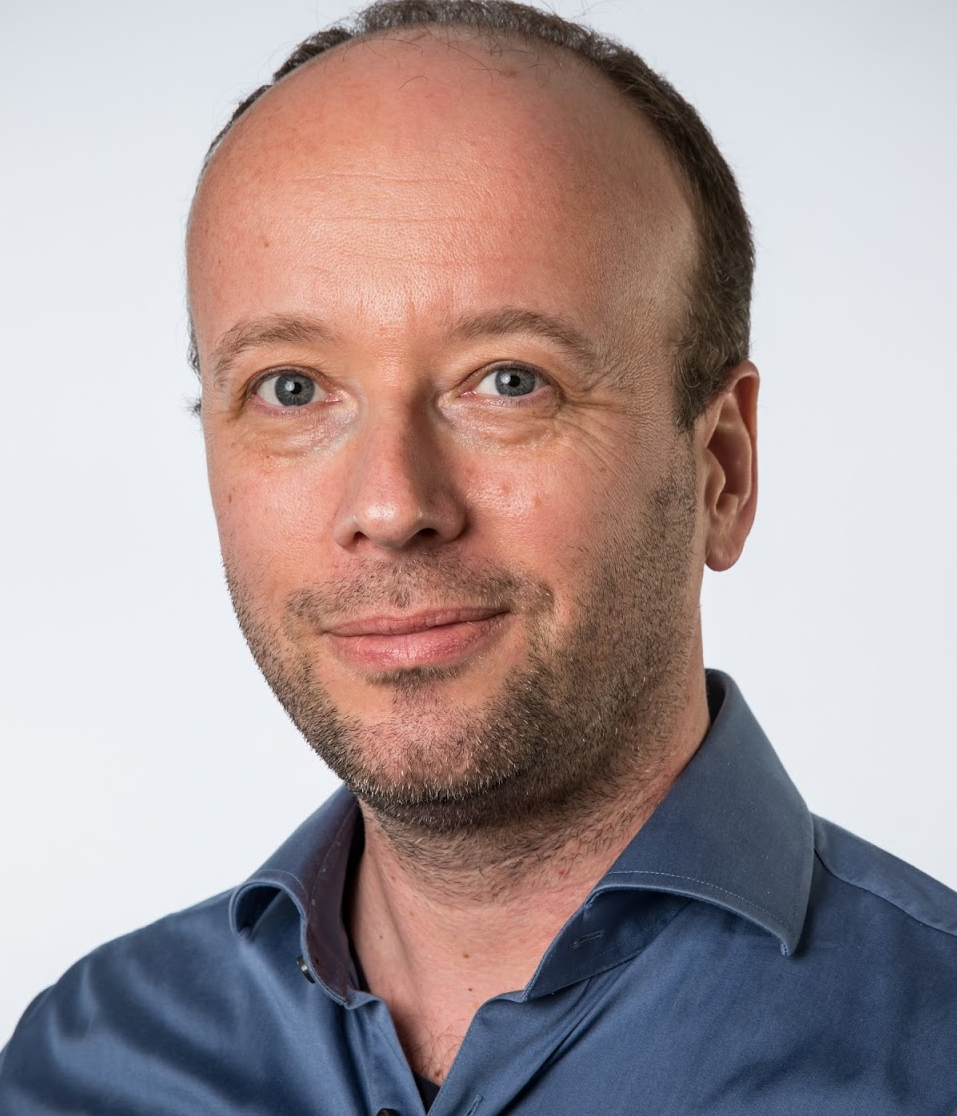}}]{Joost C. F. de Winter}
received the M.Sc. degree in Aerospace Engineering and the Ph.D. degree (cum laude) in Mechanical Engineering from the Delft University of Technology, Delft, The Netherlands, in 2004 and 2009, respectively. He is currently an Associate Professor with the Mechanical Engineering Department, Delft University of Technology. His research interests include human factors and statistical modeling, including the study of individual differences and driver behavior modeling
\end{IEEEbiography}


\end{document}